\documentclass[conference]{IEEEtran}
\IEEEoverridecommandlockouts
\usepackage{cite}
\usepackage{amsmath,amssymb,amsfonts}
\usepackage{algorithmic}
\usepackage{graphicx}
\usepackage{textcomp}
\usepackage{xcolor}

\usepackage{amsmath}
\usepackage{amsthm}
\usepackage{booktabs}
\usepackage{multirow}
\usepackage{url}
\usepackage{hyperref}

\newcommand{\iec}{i.\,e.,\,}

\newcommand{\egc}{e.\,g.,\,}
\newcommand{\eg}{e.\,g.\,}

\newcommand{\Sect}[1]{Section~\ref{#1}}

\newcommand{\Fig}[1]{Figure~\ref{#1}}

\newcommand{\Table}[1]{Table~\ref{#1}}

\def\BibTeX{{\rm B\kern-.05em{\sc i\kern-.025em b}\kern-.08em
    T\kern-.1667em\lower.7ex\hbox{E}\kern-.125emX}}
\begin{document}

\title{Multi-modal Conditional Bounding Box Regression for Music Score Following}

\author{\IEEEauthorblockN{Florian Henkel$^1$, Gerhard Widmer$^{1,2}$}
\IEEEauthorblockA{$^1$Institute of Computational Perception \& $^2$LIT Artificial Intelligence Lab \\
Johannes Kepler University Linz, Austria \\
florian.henkel@jku.at}
}
\maketitle

\begin{abstract}

This paper addresses the problem of sheet-image-based on-line audio-to-score alignment also known as score following.
Drawing inspiration from object detection, a conditional neural network architecture is proposed that directly predicts x,y coordinates of the matching positions in a complete score sheet image at each point in time for a given musical performance.

Experiments are conducted on a synthetic polyphonic piano benchmark dataset and the new method is compared to several existing approaches from the literature for sheet-image-based score following as well as an Optical Music Recognition baseline.

The proposed approach achieves new state-of-the-art results and furthermore significantly improves the alignment performance on a set of real-world piano recordings by applying Impulse Responses as a data augmentation technique. 

\end{abstract}


%%%%%%%%%%%%%%%%%%%%%%%%%%%%%%%%
%%%%%%%% Introduction %%%%%%%%%%
%%%%%%%%%%%%%%%%%%%%%%%%%%%%%%%%
\section{Introduction}

Audio-to-score alignment is a well known task in music signal processing \cite{MontecchioC11_RealtimeA2S_ICASSP, Cont06_ScoreFollowingViaNmfAndHMM_ICASSP, OtsukaNTOO11_A2SParticleFilter_EURASIP, AgrawalD20_FrameSimilarityA2S_EUSIPCO, JoderER11_CRFA2S_TASLP}, where a musical performance and a corresponding score have to be synchronized along the time axis.
This can be approached as an offline task when audio and score are available upfront, or in an on-line fashion if the complete audio is not immediately accessible but incrementally processed. This is also referred to as real-time alignment or score following.
Score following has many applications in music, such as automatic page turning for musicians \cite{ArztWD08_PageTurning_ECAI}
or synchronised information display in the concert hall \cite{ArztFGGGW15_AIConcertgebouw_IJCAI}.

While traditional score following methods usually rely on computer-readable scores\cite{Arzt16_MusicTracking_PhD, NakamuraCCOS15_ScoreFollowingSemiHMM_ISMIR, SchwarzOS04_scoreFollowing_ICMC, Dixon05_ODTW_IJCAI}, \eg MIDI or MusicXML,  recent deep learning based approaches try to avoid the need of such score representations by directly following along in a sheet image \cite{DorferAW16_ScoreFollowDNN_ISMIR, DorferHW18_ScoreFollowingAudioSheet_ISMIR, HenkelBDW19_ScoreFollowingRL_TISMIR, HenkelKW20_LearningReadAndFollow_ISMIR}.
Initially, methods falling in the latter category required the score image to be available in an unrolled form: the staves have to be detected on the page, cut out and stitched together to a long sequence.
To avoid the need for an external system for this sub-problem, \cite{HenkelKW20_LearningReadAndFollow_ISMIR} proposed to treat score following as a \textit{referring image segmentation} task, which for the first time allowed to follow musical performances in full-page sheet images without any pre-processing.

One drawback of this approach is the predicted segmentation mask, which assigns a probability to each pixel in the sheet image indicating whether it corresponds to the given audio. Apart from being counter-intuitive, since there is no clear musical correspondence between a pixel by itself and the audio, it also allows the system to predict multiple regions on the sheet image (see Figure 1 in \cite{HenkelKW20_LearningReadAndFollow_ISMIR}).

To overcome this, we propose a conditional object detection architecture based on the family of YOLO object detectors \cite{RedmonDGF16_YOLO_CVPR}, and subsequently treat score following as a bounding box regression
instead of a segmentation task. By incorporating a confidence score (to be described in \Sect{subsec:cyolo}) our system offers a clear distinction between different predicted positions on the sheet image.
We empirically show that our proposed method outperforms existing approaches on MSMD, a synthetic polyphonic piano benchmark dataset \cite{DorferHAFW18_MSMD_TISMIR}. 

A further problem in previous work is the performance gap between the synthetic audio the models are trained on compared to real-world piano recordings. To alleviate this, we show that \textit{Impulse Responses} (IRs) used for data-augmentation are a simple yet effective way to allow the network to generalize better across various recording scenarios.\footnote{Code and data will be made available on-line: \url{https://github.com/CPJKU/cyolo_score_following}}

%%%%%%%%%%%%%%%%%%%%%%%%%%%%%%%%
%%%%%%%% Related Work %%%%%%%%%%
%%%%%%%%%%%%%%%%%%%%%%%%%%%%%%%%
\section{Related Previous Work}

Dorfer et al. \cite{DorferAW16_ScoreFollowDNN_ISMIR} propose to treat score following as a \textit{one-dimensional localization task}, where a small sheet image excerpt is discretized into k bins and the most likely bin matching a given short audio excerpt has to be predicted. While our proposed approach can also be seen as a localization task, we do not rely on sheet image excerpts and discretization, but directly predict the matching position for a musical performance in the complete sheet image (see \Fig{fig:task_plot}).

In \cite{DorferHW18_ScoreFollowingAudioSheet_ISMIR} and \cite{HenkelBDW19_ScoreFollowingRL_TISMIR}, score following is formulated as a \textit{reinforcement learning problem}. Agents are trained to follow along a musical performance in an unrolled score image by adapting their reading speed, \iec how fast they want to move forward or backward in the score.

In \cite{HenkelKW19_AudioConditionedUNet_WORMS}, a conditional U-Net architecture is proposed that predicts a \textit{segmentation mask} for all matching positions in a complete score sheet image for monophonic piano music. This was later extended in \cite{HenkelKW20_LearningReadAndFollow_ISMIR} to a full score-following setup incorporating long temporal context on polyphonic piano music. As a conditioning mechanism the \textit{Feature-wise linear modulation} (FiLM) \cite{PerezSDVDC18_FILM_AAAI} layer is used, which we also apply in our proposed conditional object detection architecture.

In terms of bounding box prediction in images, current object detection systems heavily rely on deep learning and can be broadly categorized into one-stage detectors, \eg YOLO \cite{RedmonDGF16_YOLO_CVPR}, and two-stage detectors, \eg Faster R-CNN \cite{RenHGS15_FasterRCNN_NEURIPS}.
YOLO-based object detectors are known for their fast inference and real-time capabilities, which is also a requirement for on-line score following and the reason we choose it for our work.

%%%%%%%%%%%%%%%%%%%%%%%%%%%%%%%%
% Conditional Object Detection %
%%%%%%%%%%%%%%%%%%%%%%%%%%%%%%%%
\section{Conditional Object Detection for Music Score Following}

\begin{figure}[t]
\centering
\includegraphics[width=0.99\columnwidth]{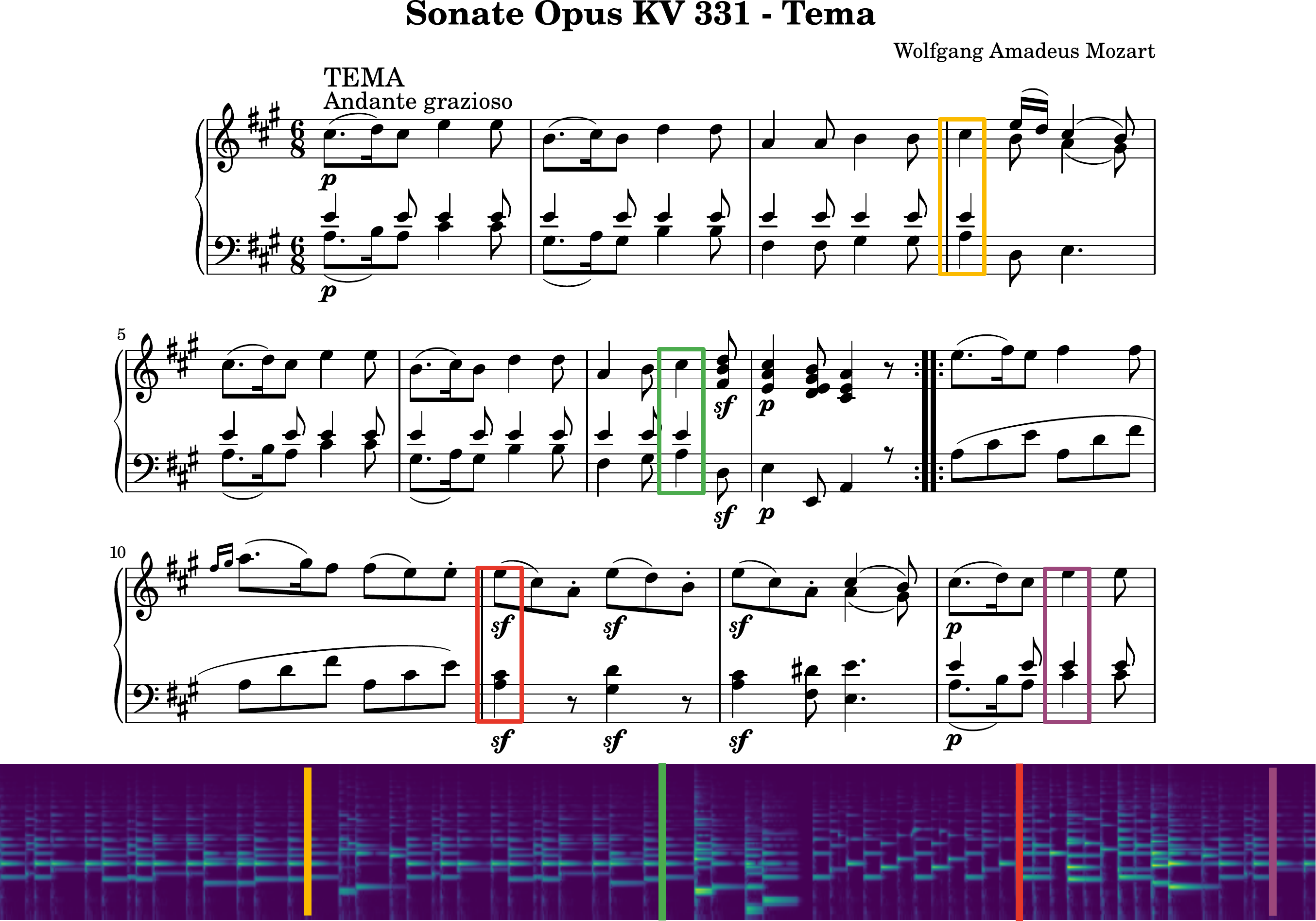}
 \caption{Score following as a bounding box regression task. For each audio frame, the model is asked to output the most likely position in the sheet image that matches the audio up until the current point in time. The colored bars in the spectrogram correspond to the bounding boxes in the sheet image.}
\label{fig:task_plot}
\end{figure}
The goal in object detection is to predict bounding boxes for all objects in an image that belong to classes the model has seen during training. In contrast, we have a \textit{conditional} setup: we want to query the model for a particular object instance, \iec instead of predicting bounding boxes for all objects in an image, the model should only predict the bounding box for a specific object based on a query.\footnote{To clarify: what is to be enclosed in a bounding box, in our case, is not an `object' proper but a position in the score sheet, with the bounding box indicating the notes or chord corresponding to the notes that have just started in the audio (see \Fig{fig:task_plot}). Desired width and height of such boxes are implicitly defined by our training data.
For the same reason, we will evaluate our score follower (in \Sect{sec:results}) by measuring the precision with which it tracks note onsets, rather than by bounding-box overlap between prediction and ground truth; after all, this is what will matter in score following.}
Similar to referring image segmentation \cite{HuRD16_ReferringSegmentation_ECCV}, this query could be some language expression referring to the desired object. In our work, instead of language expressions, we use an \textit{audio signal} as the conditioning information, similar to \cite{HenkelKW20_LearningReadAndFollow_ISMIR}.
More specifically, we will query the object detection model repeatedly for every incoming audio frame (roughly every 50ms) asking for the position in the sheet image that matches the incoming music up to the current time step (cf. \Fig{fig:task_plot}).
This is implemented as a \textit{bounding box regression task}, \iec the task of the network will be to predict the $x,y$ center-coordinates as well as the width and height of the bounding box that matches the query. In this particular work, the height depends on the height of the corresponding staff and the width is arbitrarily chosen to be 30 pixels wide in the training data annotations.

\subsection{Conditional YOLO}
\label{subsec:cyolo}
\begin{figure}[t]
\centering
\includegraphics[width=0.95\columnwidth]{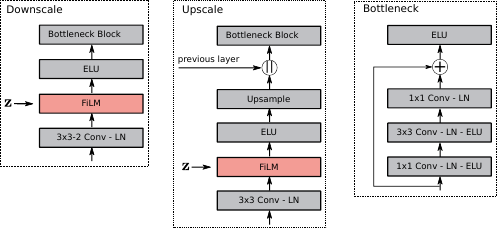}
 \caption{The main building blocks of the architecture shown in \Table{tab:architecture}. The \textit{Downscale} block applies a $3\times3$ convolution with a stride of $2$, thus halving the spatial dimension of the input feature maps. The FiLM layer in both the \textit{Downscale} and $Upscale$ block is optional and, if not indicated otherwise, simply bypassed, \iec the output of the convolutional layer is directly fed to the activation function. The \textit{Upscale} block applies nearest neighbour up-sampling by a factor of $2$ and performs a channel-wise concatenation with the feature maps from a given previous layer. The \textit{Bottleneck} block has a residual element-wise addition and reduces the number of parameters by performing the $3\times 3$ convolution with only half the number of output channels, \egc $64 \times 1 \times 1$, $64 \times 3 \times 3$ and $128 \times 1 \times 1$. If input and output do not share the same number of channels an additional $1\times 1$ convolution followed by layer normalization is applied in the residual branch before the element-wise addition.}
\label{fig:architecture}
\end{figure}

Our proposed model belongs to the family of YOLO object detectors \cite{RedmonDGF16_YOLO_CVPR} and
consists of residual \textit{Downscale} and \textit{Upscale} blocks incorporating \textit{Feature-wise linear modulation} (FiLM) \cite{PerezSDVDC18_FILM_AAAI} layers at several positions (cf. \Table{tab:architecture} and \Fig{fig:architecture}).

The FiLM layers are used to inject external information (given as a conditioning vector $\mathbf z$) into the network, by performing the following computation on feature maps $\mathbf x$
\begin{equation}
    f_{\text{FiLM}}(\mathbf{x}) = \mathbf{s}(\mathbf{z}) \cdot \mathbf{x} + \mathbf{t}(\mathbf{z}),
\end{equation}
with $\mathbf{s}(\cdot)$ and $\mathbf{t}(\cdot)$ being learned linear functions that scale and translate the input $\mathbf x$  based on $\mathbf z$. The computation of $\mathbf z$ is described in \Sect{subsec:conditioning}. For more information we refer the reader to \cite{HenkelKW20_LearningReadAndFollow_ISMIR}.

\begin{table}[t]
 \centering
 \caption{The conditional YOLO architecture consisting of several Downscale and Upscale blocks as depicted in \Fig{fig:architecture}. Upscale blocks concatenate the input from a previous layer given in parentheses, \egc layer 6 takes the output of layer 4 as additional input. FiLM indicates that the conditional layer is active in these blocks whereas it is bypassed in the others.}
 \small
 \begin{tabular}{cccc}
 \toprule
 \multicolumn{4}{c}{Conditional YOLO} \\
 \midrule
 Layer & Module & Channels &  Output Size\\
 \midrule

1  & Focus               & $16$    & $208 \times 208$\\
2  & Downscale           & $32$    & $104 \times 104$\\
3  & Downscale           & $64$    & $52 \times 52$\\
4  & Downscale - FiLM    & $128$   & $26 \times 26$\\
5  & Downscale - FiLM    & $128$   & $13 \times 13$\\
6  & Upscale(4) - FiLM   & $128$   & $26 \times 26$\\
7  & Upscale(3) - FiLM   & $128$   & $52 \times 52$\\
8  & Detection           & $15$    & $52 \times 52$\\

\midrule
\end{tabular}

 \label{tab:architecture}
\end{table}

As shown in \Table{tab:architecture}, the first layer in the network is a so-called \textit{Focus} or \textit{SpaceToDepth} \cite{RidnikLNF20_TResNet_Arxiv} layer which splits a $c\times w\times h$ image (in our case of shape $1 \times 416 \times 416$) into a $4*c\times \frac{w}{2}\times \frac{h}{2}$ ($4 \times 208 \times 208$) dimensional image with the purpose of reducing the image size and improving the overall computation speed. This is followed directly by a $16\times 3\times 3$ convolution, layer normalization\cite{BAKH16_LayerNorm_arxiv} and ELU activation\cite{ClevertUH15_ELU_ICLR}.
Each \textit{Downscale} block reduces the input size by half, thus creating feature maps of different resolutions ($104 \times 104$, $52 \times 52$, $26 \times 26$, $13 \times 13$). In the \textit{Upscale} blocks, these feature maps are up-sampled, concatenated, and subsequently used in the \textit{Detection} layer to create the final predictions.

The predictions are partly based on anchor points for which the \textit{Detection} layer predicts offsets to produce the final width and height of the bounding box. Given that we consider uniform objects of a fixed width and similar height that depend on the staves, we only use three anchor points $(11, 26),  (11,34),  (11,45)$.
$11$ indicates the anchor-width and $26, 34, 45$ the anchor-heights for a down-scaled sheet image (cf. \Sect{sec:experiments}). 
These are determined using k-means clustering of the training set bounding box \cite{RedmonF17_YOLO9000_CVPR}.

For each spatial position of the $52\times 52$ output grid and for each anchor point, the \textit{Detection} layer produces 5 outputs. The first two are offsets for the $x,y$ center coordinates of the bounding boxes. Based on these and the spatial position in the grid, the coordinates in the original input image will be computed. The next two outputs correspond to the aforementioned width and height offsets and the last output is an objectness or confidence score predicting the \textit{Intersection over Union} (IoU) for a bounding box. Colloquially, this tells us how well the predicted box fits the ground truth object. During inference, this score is used to filter the most likely position in the sheet image that matches the given audio. 

\subsection{Audio Encoding}
\label{subsec:conditioning}

The audio waveform is sampled with $22.05$ kHz and subjected to an STFT with a Hann window  of size $2048$ and a hop size of $1102$ (approx. $20$ frames per second). Each frame is further transformed using a logarithmic filterbank processing frequencies between $60$ Hz and $6$ kHz. The resulting output has 78 log-frequency bins.
Given the on-the-fly data augmentation during training to be described in \Sect{subsec:IR}, we cannot normalize the input spectrograms to zero mean and unit variance beforehand as in \cite{HenkelKW20_LearningReadAndFollow_ISMIR}. To still allow for input normalization, we add a batch normalization layer \cite{IoffeS15_BatchNorm_arxiv} in front of our encoding network that computes mean and standard deviation parameters for each frequency band similar to \cite{GrillS17_BirdDetection_EUSIPCO}.
To encode the spectrogram, we apply the same CNN encoder as in \cite{HenkelKW20_LearningReadAndFollow_ISMIR}, projecting 40 audio frames to a 32 dimensional vector $\mathbf{x}$. To encode the larger temporal audio context, we apply an LSTM layer with 64 hidden units \cite{HochreiterS97_LSTM_NeuralComp}. Its hidden state is updated every 40 audio frames using $\mathbf{x}$. The final  conditioning vector $\mathbf{z}$ that will be used in the FiLM layers is computed as 
\begin{equation}
    \mathbf{z} = f([\mathbf{h};\mathbf{x}]),
\end{equation}
where $\mathbf{h}$ is the 64 dimensional hidden state vector of the LSTM, $[\mathbf{h};\mathbf{x}]$ indicates concatenation of $\mathbf{h}$ and $\mathbf{x}$, and $f$ is a fully connected layer of size 128 with layer normalization and ELU activation. Thus, the final dimensionality of $\mathbf{z}$ is 128. 

%%%%%%%%%%%%%%%%%%%%%%%%%%%%%%%%
%%%%%%%%% Experiments %%%%%%%%%%
%%%%%%%%%%%%%%%%%%%%%%%%%%%%%%%%
\section{Experimental Setup}
\label{sec:experiments}
We conduct our experiments using the Multi-modal Sheet Music Dataset (MSMD) \cite{DorferHAFW18_MSMD_TISMIR}, a piano music dataset which offers alignments between notehead positions in the sheet image and a synthetic MIDI score. For training, this MIDI is rendered to an audio performance using a publicly available piano sound-font with Fluidsynth.\footnote{\url{https://www.fluidsynth.org}}
As in \cite{HenkelKW20_LearningReadAndFollow_ISMIR}, the dataset is split into 945 train, 28 validation and 125 test pages. The sheet image pages with an initial shape of $1181\times835$ are zero-padded to a squared size and down-scaled to $416\times 416$. This still offers a high enough resolution such that humans are able to detect details in the sheet image.

The \textit{optimization objective} consists of two parts, a mean-squared-error loss for the bounding boxes as well as a logistic regression loss for predicting the aforementioned objectness score, which should reflect the IoU score of a predicted bounding box, \iec a predicted box with a perfect overlap to the ground truth should have a score of $1$\cite{RedmonF17_YOLO9000_CVPR}.

For training, we use the Adam optimizer with corrected weight decay regularization \cite{LoshchilovH19_WeightDecayAdam_ICLR}, and a weight decay coefficient of $5e^{-4}$. Weight decay is only applied to the weight parameters of convolutional, recurrent and linear layers, but not to normalization layers and bias parameters \cite{HeZZZXL19_BagOfTricks_CVPR}. 
The weights of the network are initialized orthogonally \cite{SaxeMG13_OrthogonalInit_arxiv} and the biases are set to zero except for the forget gate bias in the LSTM layer which is set to $1$\cite{GersSC00_LearningTOForget_NeuralComp}. 
We anneal the learning rate over the course of 50 epochs starting from $5e^{-4}$ to $5e^{-6}$ using a cosine annealing scheme\cite{LoshchilovH17_SGDR_ICLR}. 
The best models in terms of the validation loss are selected for evaluation on the test set.
To avoid exploding gradients, we apply gradient clipping with a maximum norm of $0.1$ to the parameters of the recurrent layer in the audio encoder.

We perform data augmentation in the sheet image domain by randomly shifting the score along the $x,y$ axis and additionally augment the audio by changing the tempo of the performance with a random factor between $0.5$ and $2$.

\subsection{Impulse Response Data Augmentation}
\label{subsec:IR}
One issue with existing approaches to sheet-image-based score following is their lack of generalization to real-world audio \cite{HenkelBDW19_ScoreFollowingRL_TISMIR,HenkelKW20_LearningReadAndFollow_ISMIR}. This is attributed to the synthetic data these models are trained on, due to the shortage of alignments between real-world audio recordings and notehead positions in sheet music.
To alleviate this, we apply \textit{Impulse Responses} (IRs) as an audio data augmentation technique. IRs allow us to model different recording conditions in the form of microphones and room characteristics. We use more than $500$ freely available IRs collected from OpenAIRLib\footnote{\url{https://openairlib.net}} and MicIRP.\footnote{\url{https://micirp.blogspot.com}}
The IR signal is convolved with the audio on-the-fly during training, such that in each epoch the model encounters different audio scenarios, which should allow for a more robust audio encoding model. To measure the effect of this technique, we train our model with and without IR augmentation.

\subsection{Baselines}
\label{subsec:baselines}
As baselines we choose three methods for sheet-image-based score following: the supervised localization network referred to as MM-Loc \cite{DorferAW16_ScoreFollowDNN_ISMIR}, the best performing reinforcement learning (RL) agent from \cite{HenkelBDW19_ScoreFollowingRL_TISMIR}, and the conditional segmentation network \cite{HenkelKW20_LearningReadAndFollow_ISMIR}, which we refer to as CUNet. Additionally, a more traditional method based on optical music recognition (OMR) is considered that first performs OMR on a given sheet image to create a MIDI score representation. The MIDI is then rendered to audio and on-line dynamic time warping (ODTW)\cite{Mueller15_FMP_SPRINGER} is applied to perform audio-to-audio alignment.

%%%%%%%%%%%%%%%%%%%%%%%%%%%%%%%%
%%%%%%%%%%% Results %%%%%%%%%%%%
%%%%%%%%%%%%%%%%%%%%%%%%%%%%%%%%
\section{Results}
\label{sec:results}

\Table{tab:results} summarizes the results across different settings. As an evaluation metric, we choose the ratio of tracked onsets below certain error thresholds ranging from $0.05$ to $5$ seconds similar to \cite{HenkelKW20_LearningReadAndFollow_ISMIR}.\footnote{Note that eventually, realistic evaluation criteria should reflect the targeted application. While accompaniment systems require very precise results (\iec low error thresholds), a page turner can get away with lower accuracy.}
To arrive at this metric, we transform the predicted bounding boxes in the sheet image back to the corresponding timestep position in the audio using the ground truth annotations. The absolute difference between the predicted timestep (transformed from the bounding box) and the ground truth note onset then yields the tracking error.
In the first section of the table, we evaluate all approaches on the full MSMD test split consisting of 125 sheet-image pages. This is a completely synthetic setup, meaning that the audio is rendered from MIDI with the same piano sound-font seen during training. We observe that our non IR-augmented model (CYOLO) already outperforms all previous methods across all error thresholds. Additionally incorporating IR-augmentation (CYOLO-IR) slightly improves the tracking performance, potentially by making it harder to overfit to the training data.
As a notable difference to the referring image segmentation method (CUNet), our model not only tracks notes more accurately but is overall able to track more notes below the $5$ second error threshold. We attribute this to our task formulation that allows us to directly predict positions in the sheet image instead of inferring them from segmented regions.

\begin{table}[t]
\centering
\caption{Comparison of our proposed methods to several approaches as described in \Sect{sec:experiments}. We report the ratio of tracked onsets below certain error thresholds from $0.05$ to $5$ seconds. The best result for each threshold is marked bold.}
\small
\begin{tabular}{lccccc}
 \toprule
\textbf{Err. [sec]} & $\leq 0.05$ &  $\leq 0.10$ & $\leq 0.50$ & $\leq 1.00$ & $\leq 5.00$\\\midrule\midrule
 \multicolumn{6}{l}{I\ \ MSMD Full Test Set (Synthetic)} \\
 \midrule
    OMR\cite{HenkelBDW19_ScoreFollowingRL_TISMIR}       & 0.447 & 0.519 & 0.760 & 0.850 & 0.974 \\
    MM-Loc \cite{DorferAW16_ScoreFollowDNN_ISMIR}       & 0.446 & 0.492 & 0.822 & 0.860 & 0.920 \\
    RL\cite{HenkelBDW19_ScoreFollowingRL_TISMIR}        & 0.409 & 0.433 & 0.797 & 0.878 & 0.972 \\
    CUNet\cite{HenkelKW20_LearningReadAndFollow_ISMIR}  & 0.733 & 0.747 & 0.852 & 0.885 & 0.937 \\
    CYOLO                                               & 0.815 & 0.826 & 0.865 & 0.890 & 0.979 \\
    CYOLO-IR                                            & \textbf{0.830} & \textbf{0.842} & \textbf{0.885} & \textbf{0.909} & \textbf{0.984} \\
  \midrule
    \midrule
 \multicolumn{6}{l}{II\ \ MSMD Test Subset (Synthetic)} \\
 \midrule
    OMR\cite{HenkelBDW19_ScoreFollowingRL_TISMIR}       & 0.371 & 0.461 & 0.749 & 0.868 & \textbf{0.996} \\
    MM-Loc \cite{DorferAW16_ScoreFollowDNN_ISMIR}       & 0.416 & 0.442 & 0.776 & 0.799 & 0.903 \\
    RL\cite{HenkelBDW19_ScoreFollowingRL_TISMIR}        & 0.365 & 0.382 & 0.729 & 0.798 & 0.965 \\
    CUNet\cite{HenkelKW20_LearningReadAndFollow_ISMIR}  & 0.698 & 0.706 & 0.806 & 0.824 & 0.891 \\
    CYOLO                                               & 0.768 & 0.777 & 0.829 & 0.846 & 0.979 \\
    CYOLO-IR                                            & \textbf{0.795} & \textbf{0.803} & \textbf{0.863} & \textbf{0.880} & 0.990 \\

 \midrule\midrule
 \multicolumn{6}{l}{III (a) \ \ Performance MIDI Synthesized} \\
 \midrule
    OMR\cite{HenkelBDW19_ScoreFollowingRL_TISMIR}       & 0.289 & 0.398 & 0.717 & 0.834 & \textbf{0.988} \\
    MM-Loc \cite{DorferAW16_ScoreFollowDNN_ISMIR}       & 0.472 & 0.490 & 0.832 & 0.861 & 0.960 \\
    RL\cite{HenkelBDW19_ScoreFollowingRL_TISMIR}        & 0.234 & 0.248 & 0.545 & 0.640 & 0.812 \\
    CUNet\cite{HenkelKW20_LearningReadAndFollow_ISMIR}  & 0.565 & 0.581 & 0.809 & 0.844 & 0.901 \\
    CYOLO                                               & 0.661 & 0.676 & 0.804 & 0.840 & 0.962 \\
    CYOLO-IR                                            & \textbf{0.759} & \textbf{0.774} & \textbf{0.856} & \textbf{0.882} & 0.987 \\
 \midrule\midrule
 \multicolumn{6}{l}{III (b) \ \ Direct Out} \\
 \midrule
    OMR\cite{HenkelBDW19_ScoreFollowingRL_TISMIR}       & 0.226 & 0.330 & 0.703 & \textbf{0.839} & \textbf{0.993} \\
    MM-Loc \cite{DorferAW16_ScoreFollowDNN_ISMIR}       & 0.338 & 0.354 & 0.597 & 0.634 & 0.753 \\
    RL\cite{HenkelBDW19_ScoreFollowingRL_TISMIR}        & 0.277 & 0.291 & 0.607 & 0.733 & 0.955 \\
    CUNet\cite{HenkelKW20_LearningReadAndFollow_ISMIR}  & 0.400 & 0.416 & 0.642 & 0.693 & 0.811 \\
    CYOLO                                               & 0.584 & 0.599 & 0.769 & 0.817 & 0.973 \\
    CYOLO-IR                                            & \textbf{0.642} & \textbf{0.656} & \textbf{0.772} & 0.806 & 0.965 \\
  \midrule\midrule
 \multicolumn{6}{l}{III (c) \ \ Room Recording} \\
 \midrule
    OMR\cite{HenkelBDW19_ScoreFollowingRL_TISMIR}      & 0.226 & 0.322 & 0.702 & \textbf{0.827} & \textbf{0.974} \\
    MM-Loc \cite{DorferAW16_ScoreFollowDNN_ISMIR}      & 0.207 & 0.243 & 0.541 & 0.573 & 0.702 \\
    RL\cite{HenkelBDW19_ScoreFollowingRL_TISMIR}       & 0.192 & 0.206 & 0.466 & 0.587 & 0.891 \\
    CUNet\cite{HenkelKW20_LearningReadAndFollow_ISMIR} & 0.094 & 0.105 & 0.215 & 0.262 & 0.443 \\
    CYOLO                                              & 0.306 & 0.323 & 0.460 & 0.510 & 0.763 \\
    CYOLO-IR                                           & \textbf{0.563} & \textbf{0.581} & \textbf{0.712} & 0.749 & 0.919 \\
 \bottomrule
 \end{tabular}
 \label{tab:results}
\end{table}

Starting with section II of the table, we now limit the evaluation to a subset of 25 pages from the MSMD test set for which we also have real piano recordings that were aligned to the sheet images. This will allow us to measure how much expressive performance deviations, played ornaments, errors, and different audio conditions affect the score followers. As a baseline, section II shows the results on the synthetic audios for just this subset of pieces.

In the remaining three sections (III a-c), we have the same 25 pages from the test set, with (a) the recorded \textit{performance} MIDI synthesized by the sound-font used during training, (b) the audio recorded from the direct output of a Yamaha AvantGrand N2 hybrid piano, and (c) the audio captured with a microphone in an office room. 

(a) allows us to observe the effect of expressive playing in contrast to a deadpan MIDI performance, while still having a similar audio condition as during training.
Indeed, we see some slight deterioration relative to II, overall.
Interestingly, even though we are still in the synthesized piano sound domain, the IR audio augmentation again seems to help -- this is likely due to the loudness variations that come with the real performances but did not appear in the original synthesized audios.

In (b) we see the first results for different playing as well as audio conditions. Again, our method performs best across the three lowest error thresholds and using IR-augmentation further improves the results. Apart from the OMR approach, we outperform all other sheet-image-based methods.

(c) offers the most difficult challenge as the audio conditions are particularly different to what was seen during training, due to having been recorded in an office environment, including background noise as well as reverberation effects from the room.
In this setting we clearly observe how the IR-augmentation considerably improves the results. Our method outperforms the sheet-image-based approaches across all, and the OMR-based approach for the three lowest error thresholds.
There is still a performance gap to the strictly synthetic setting, however our method offers a big step towards closing it. We assume that without the use of additional (real-world) data we will probably not be able to further reduce this gap.

%%%%%%%%%%%%%%%%%%%%%%%%%%%%%%%%
%%%%%%%%% Conclusion %%%%%%%%%%%
%%%%%%%%%%%%%%%%%%%%%%%%%%%%%%%%
\section{Conclusion}
We proposed a new architecture for sheet-image-based score following, inspired by an object detection setup. The method compares favorably to existing methods from the literature and significantly improves the performance for real-world piano recordings.
Besides, our new approach also promises to be easily extendable. In future work, we plan to not only predict the most likely note-level position, but also the corresponding bar and system.
We argue, that this could lead to additional stability for our network and furthermore allow us to leverage more data, where no fine-granular note-level alignment between audio and sheet music is available. This would make it easier to use scanned or photographed scores, and thus make it possible to improve and evaluate the generalization capabilities not only in the audio but also in the sheet-image domain.

\section*{Acknowledgment}

 This work was supported by the European Research Council (ERC) under the EU's Horizon 2020 research and innovation program (grant number 670035 "Con Espressione").
 
\bibliographystyle{IEEEtran}
\bibliography{main}

\end{document}